\documentclass[twocolumn, times]{aastex61}
\usepackage{amsmath}
\usepackage{lineno}

\hypersetup{colorlinks=true, breaklinks, linkcolor=blue, citecolor=blue, filecolor=blue, urlcolor=blue}

\shorttitle{IRAS16316-1540}
\shortauthors{Yoon et al.}

\begin{document}

\title{Evidence of Accretion Burst: The Viscously Heated Inner Disk of the Embedded Protostar IRAS 16316-1540}

\accepted{June 29, 2021 to ApJ}
\correspondingauthor{Jeong-Eun Lee}
\email{jeongeun.lee@khu.ac.kr}

\author{Sung-Yong Yoon}
\affiliation{The School of Space Research, Kyung Hee University, 1732 Deogyeong-daero, Giheung-gu, Yongin-si, Gyeonggi-do, Republic of Korea\\ :\href{mailto:jeongeun.lee@khu.ac.kr}{jeongeun.lee@khu.ac.kr}}
\affiliation{Korea Astronomy and Space Science Institute, 776, Daedeok-daero, Yuseong-gu, Daejeon, 34055, Republic of Korea\\ :\href{mailto:yoonsy@kasi.re.kr}{yoonsy@kasi.re.kr}}

\author[0000-0003-3119-2087]{Jeong-Eun Lee}
\affiliation{The School of Space Research, Kyung Hee University, 1732 Deogyeong-daero, Giheung-gu, Yongin-si, Gyeonggi-do, Republic of Korea\\ :\href{mailto:jeongeun.lee@khu.ac.kr}{jeongeun.lee@khu.ac.kr}}

\author{Seokho Lee}
\affiliation{National Astronomical Observatory of Japan, 2-21-1 Osawa, Mitaka, Tokyo 181-8588, Japan}

\author{Gregory J. Herczeg}
\affiliation{Kavli Institute for Astronomy and Astrophysics, Peking University, Yiheyuan Lu 5, Haidian Qu, 100871 Beijing, China}
\affiliation{Department of Astronomy, Peking University, Yiheyuan 5, Haidian Qu, 100871 Beijing, China}

\author[0000-0003-4099-1171]{Sunkyung Park}
\affiliation{Konkoly Observatory, Research Centre for Astronomy and Earth Sciences, E\"otv\"os Lor\'and Research Network (ELKH), Konkoly-Thege Mikl\'os \'ut 15-17, 1121 Budapest, Hungary}

\author{Gregory N. Mace}
\affiliation{Department of Astronomy, University of Texas at Austin, 2515 Speedway, Austin, TX, USA}

\author{Jae-Joon Lee}
\affiliation{Korea Astronomy and Space Science Institute, 776, Daedeok-daero, Yuseong-gu, Daejeon, 34055, Republic of Korea}

\author{Daniel T. Jaffe}
\affiliation{Department of Astronomy, University of Texas at Austin, 2515 Speedway, Austin, TX, USA}

\begin{abstract}

Outbursts of young stellar objects occur when the mass accretion rate suddenly increases.  However, such outbursts are difficult to detect for deeply embedded protostars due to their thick envelope and the rarity of outbursts. The near-IR spectroscopy is a useful tool to identify ongoing outburst candidates by the characteristic absorption features that indicate a disk origin. However, without high-resolution spectroscopy, the spectra of outburst candidates can be confused with the late-type stars since they have similar spectral features. For the protostar IRAS 16316-1540, the near-IR spectrum has line equivalent widths that are consistent with M-dwarf photospheres. However, our high-resolution IGRINS spectra reveal that the absorption lines have boxy and/or double-peaked profiles, as expected from a disk and not the star. The continuum emission source is likely the hot, optically thick disk, heated by viscous accretion.  The projected disk rotation velocity of 41$\pm$5 km s$^{-1}$ corresponds to $\sim 0.1$ AU.  Based on the result, we suggest IRAS 16316-1540 as an ongoing outburst candidate. Viscous heating of disks is usually interpreted as evidence for ongoing bursts, which may be more common than previously estimated from low-resolution near-IR spectra.

\end{abstract}

\section{Introduction} \label{sec:intro}

How protostars grow in mass is a long-standing question. In low-mass star formation, material is transferred from an infalling envelope to a disk, then accretes onto a protostar. However, it is still unclear how mass is transported inward through the disk and into the protostar. The standard model of star formation \citep{Shu77, Terebey84} predicts that the mass accretion onto the protostar proceeds at a constant rate. However, accretion luminosities estimated from constant accretion rates are higher than the observed luminosities for embedded protostars \citep[known as the luminosity problem;][]{Kenyon90, Evans09}. A promising solution to this problem is the combination of long periods of quiescence interspersed with short, rare outbursts, known as episodic accretion \citep[][see also reviews by \citeauthor{Hartmann96} \citeyear{Hartmann96},  \citeauthor{Audard14} \citeyear{Audard14}]{Kenyon90, Kenyon95}. 
The repeated accretion outbursts may be explained by disk instabilities, including thermal instabilities in the inner disk \citep{Bell94}, disk fragmentation and gravitational instabilities \citep{Vorobyov15}, the combination of gravitational instability and magnetorotational instabilities \citep{Zhu09b}, and tidal interaction by companions \citep{Bonnell92}.

The existence of large accretion outbursts has been demonstrated by the discovery and characterization of FU Orionis-type objects (FUors). FUors brighten in luminosity by several magnitudes, interpreted as an increase in the accretion rate, and can remain bright for many decades \citep{Audard14}.  
However, the frequency and duration of such accretion bursts are still uncertain because of the scarcity of samples \citep{Hillenbrand15, ContrerasPena19}, and the rarity of studies that could find and characterize outbursts of protostars that may last for decades \citep{scholz13,Fischer17,Fischer19, ContrerasPena17,Guo20}. Outbursts have been detected in deeply embedded protostars (Class 0 and I) only in few cases due to their thick envelopes \citep[e.g.][]{Hodapp96,Kospal07,Safron15, Yoo17}. Since protostars gain a significant fraction of their mass in the early embedded evolutionary phase \citep[e.g.][]{Bae14}, more frequent bursts may occur at these early evolutionary phases.

While the change from quiescence to outburst is challenging to detect, especially for embedded protostars, the distinct spectral characteristics of FUors provide us with diagnostics to identify protostars with ongoing accretion bursts.
Prominent CO overtone and water vapor absorption bands have been used to identify FUor candidates \citep{Carr87, Kenyon93a, Greene97, Reipurth97, Sandell98, Aspin03, Greene08, Connelley18}. However, CO overtone and water vapor absorption bands observed at low-resolution are often insufficient to distinguish between viscously heated disks and late-type stellar photospheres because late-type stellar photospheric spectra show similar absorption features \citep{Shiba93, Greene96, Greene97, Connelley18}. 
In some cases, individual lines have double-peaked or boxy profiles that are best interpreted as disk rotation \citep{Hartmann96,Lee15,Park20}. For FUors, these absorption lines are produced by a warm inner disk of T$\rm_{eff}$$\sim$3000-4000 K, heated by viscous accretion.

In this paper, we analyze spectral features of the protostar IRAS 16316-1540 in high-resolution $H$- and $K$-band IGRINS spectra. The high resolution spectra are similar to those of FUors, indicating that IRAS 16316-1540 is likely undergoing an outburst, even though the change from quiescence to burst was not observed. In contrast, based on analyses from low-resolution IR spectra, IRAS 16316-1540 had been classified as a stellar photosphere rather than an FUor object because of the modest depths of the absorption lines \citep{Connelley10}.

IRAS 16316-1540 (also known as RNO 91 and HBC 650) is embedded in a reflection nebula with an infrared excess \citep{Herbst81, Hodapp94, Mayama07, Chen09}, located in the L43 dark cloud at a distance of 125 pc \citep{deGeus90}. SED fits yield a bolometric luminosity of 2.5 L$_{\sun}$, a bolometric temperature of 340 K, and a classification as a Class I object \citep{Chen09, kristensen12}. The low-resolution optical spectrum is heavily veiled and lacks broadband features typical of M-dwarfs but may have subtle features consistent with an early K star \citep{Herczeg14}. Optical and near-infrared (NIR) images reveal that the protostar is surrounded by a complex structure of lobes and falling knots, indicative of the interaction between the outflow and the dissipating envelope \citep{Schild89, Mayama07}. Polarization observations in the optical and NIR have suggested a disk-like structure in the nebula \citep{Scarrott93, Weintraub94,Mayama07}. High resolution $L$-band imaging ($0\farcs08$ resolution) did not reveal any companion \citep{connelley09}.

We present here the NIR spectra of IRAS 16316-1540, analyze the spectroscopically resolved absorption line profiles to investigate the origin of spectral features, and discuss the physical properties of IRAS 16316-1540. Observations and data reduction are presented in Section \ref{sec:obs}. The spectral features of IRAS 16316-1540 are described in Section \ref{sec:Spec16316}. Analysis and Discussion are given in Section \ref{sec:LineProfile} and \ref{sec:Discuss}. We summarize our results in Section~\ref{sec:final}.

\begin{deluxetable*}{ccCccccc}[t]
\tablecaption{Observation log of IRAS 16316-1540 \label{tab:log}}
\tablecolumns{8}
\tablewidth{0pt}
\tablehead{
\colhead{Observatory} &
\colhead{UT date} &
\colhead{Total exposure time} &
\colhead{Airmass} & \colhead{A0 Star} & \colhead{Air mass} & \colhead{Seeing} &
\colhead{Median SNR}\\
\colhead{/Telescope} & \colhead{(YYYY-mm-dd)} & \colhead{(sec)} &
\colhead{(IRAS 16316-1540)} & \colhead{} & \colhead{(A0 star)} & 
\colhead{(arcsec)} & \colhead{in K-band}
}
\startdata
McDonald/HJST & 2015-04-30 & 2400 & 1.476 & HIP 80974 & 1.385 & 1.92 & 110\\
McDonald/HJST & 2016-04-28 & 1800 & 1.674 & HIP 80974 & 1.779 & 2.30 & 101\\
McDonald/HJST & 2017-03-12 & 2400 & 1.653 & HD 163336 & 1.644 & 2.04 & 137\\
McDonald/HJST & 2017-04-13 & 2400 & 1.509 & HD 159008 & 1.517 & 1.65 & 138\\
McDonald/HJST & 2017-06-02 & 2400 & 1.570 & HR 5332 & 1.567 & 2.43 & 110\\
McDonald/HJST & 2017-08-03 & 2100 & 1.463 & HD 163336 & 1.470 & 1.68 & 139\\
Lowell/DCT & 2017-09-03 & 2400 & 2.085 & 50 Lib & 2.520 & 1.40 & 93\\
Gemini/GST & 2018-06-02 & 1440 & 1.046 & HIP 79229 & 1.009 & 0.74 & 474\\
\enddata
\end{deluxetable*}

\section{Observations and Data Reduction} \label{sec:obs}

We obtained the NIR spectra of IRAS 16316-1540 using Immersion Grating Infrared Spectrograph (IGRINS). IGRINS is a cross-dispersed spectrograph that obtains simultaneous $H$- and $K$-band spectra with a resolution $\text{R}=\frac{\lambda}{\Delta\lambda}\sim 45,000$.  We observed IRAS 16316-1540 at the 2.7 m Harlan J. Smith telescope (HJST) of the McDonald Observatory from 2015 April 30 to 2017 August 3, the 4.3 m Discovery Channel Telescope (DCT) of the Lowell Observatory on 2017 September 3, and the Gemini-South Telescope (GST) of the Gemini Observatory on 2018 June 2 (see observation log in Table~\ref{tab:log}). Spectra were obtained with slit scales of 1\arcsec\,$\times$\,15\arcsec, 0.63\arcsec\,$\times$\,9.3\arcsec, and 0.34\arcsec\,$\times$\,5\arcsec~for HJST, DCT, and GST, respectively. The seeing for each observation, as listed in Table~\ref{tab:log}, is estimated by measuring the width of the spectrum in the cross-dispersion direction.  Most observations were conducted in a sequence of ABBA nods, in which the target position is moved from side to side along the slit to subtract dark current, sky, and telescope backgrounds \citep{Cushing04}. The data taken on 2017 August 3 was obtained by nodding between the object and a blank sky position.

The data were reduced using the IGRINS pipeline \citep{Lee17}. The raw data were flat-fielded to correct the pixel-to-pixel variations. Bad pixels were masked from flat-off and flat-on images. A modified version of the optimal extraction algorithms of \citet{Horne86} was used to extract one-dimensional spectra. An A0 V standard star in the vicinity of the target was observed just after or before each target observation to remove the telluric absorption lines. The OH lines of blank sky and the telluric absorption lines of the A0 V star were used to derive wavelength solutions. The A0 V spectrum was divided by a Vega model spectrum to remove hydrogen recombination absorption lines that appear in the A0 V spectrum. We scaled the A0 V spectrum to remove the effect from the difference in airmass between the target and A0 V star. By dividing the target spectrum by the scaled A0 V spectrum, the telluric lines in the target spectrum were removed. We normalized the reduced spectrum with the {\tt\string continuum} task of IRAF\footnote{IRAF is the Image Reduction and Analysis Facility, a general purpose software system for the reduction and analysis of astronomical data. \url{http://iraf.noao.edu/}}.

\begin{figure*}[ht!]
    \includegraphics[width=0.99\textwidth]{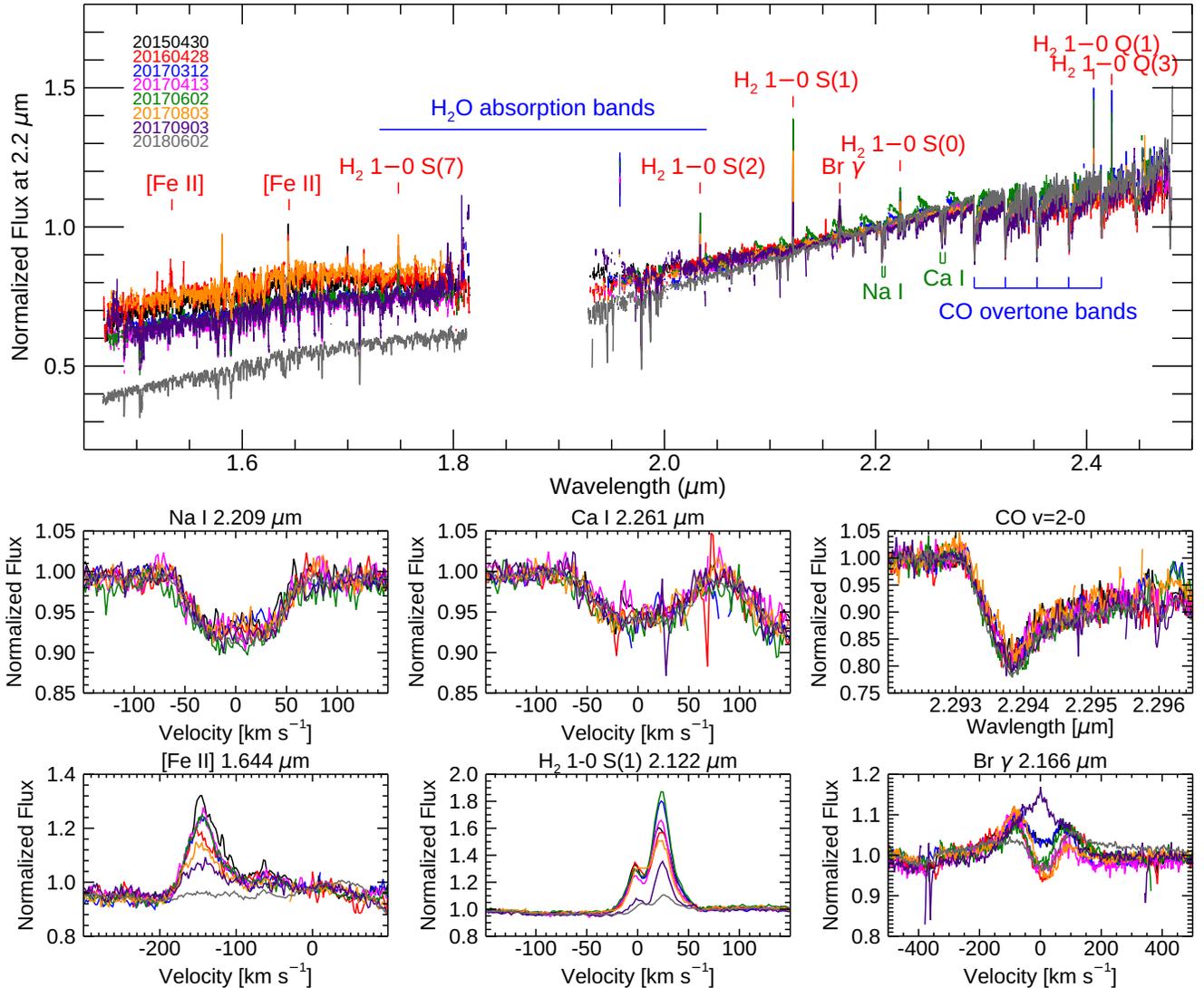}
    \caption{IGRINS spectra of IRAS 16316-1540 in the H- and K-bands. The spectral lines of emission and absorption are presented in each panel. The spectra of each observations epoch are shown in other colors which are referred in the top panel. The absorption spectra are consistent over the observation epochs, while emission lines show variation in strength and profile, for most lines due to differences in amount of extended emission included in the spectrum. \label{fig:HKband}}
\end{figure*}

\section{Spectral features of IRAS 16316-1540} \label{sec:Spec16316}

We present the NIR spectra of IRAS 16316-1540 in Figure~\ref{fig:HKband}, obtained in eight epochs across from 2015--2018.
The H$_2$O absorption band from 1.7 to 2.0 $\mu$m makes the H-band continuum a curved or triangular shape. CO overtone absorption bands appear in K-band spectra starting at 2.29 $\mu$m.
The NIR spectra also show many other atomic and molecular absorption lines, including Na I and Ca I, that are typically prominent in the late-type stars and YSOs, not in FUors \citep{Greene96, Greene97, Wallace97}.  Br$\gamma$ 2.166 $\mu$m, [\ion{Fe}{2}] 1.533, 1.644 $\mu$m and several molecular hydrogen (H$_2$) lines are seen in emission. Since the absorption line profiles do not show any variability between the observation epochs, all spectra were weighted by observation errors and co-added prior to measurements.  The emission lines vary between epochs.

\begin{figure*}[ht!]
\centering
\includegraphics[width=0.99\textwidth]{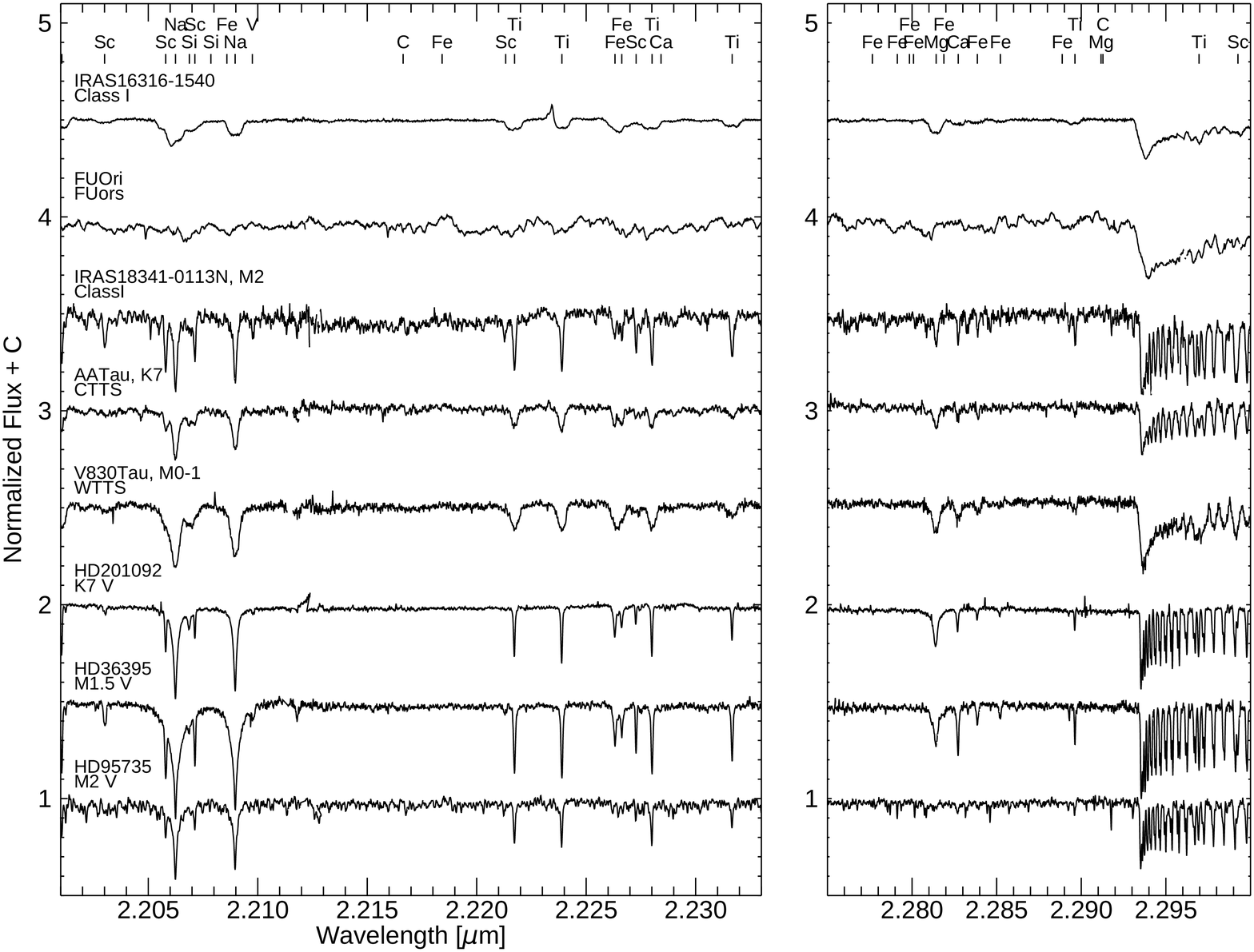}
\caption{The NIR spectra of IRAS 16316-1540, several types of YSOs, and dwarfs. The spectral regions include Na I doublet and CO overtone bandhead. The positions of spectral lines are denoted on top of the spectra. The absorption spectral features in the dwarfs are well traced in YSOs and IRAS 16316-1540, while the spectra of FU Ori are heavily blended so obscure other than the CO overtone band. \label{fig:wOthers}}
\end{figure*}

Figure~\ref{fig:wOthers} compares the spectra of IRAS 16316-1540 with an FUor, YSOs, and dwarf stars from the IGRINS spectral library \citep{Park18}. All of the objects in Figure~\ref{fig:wOthers} except FU Ori show prominent neutral atomic absorption lines including Na, Ti, Mg, and Fe. FU Ori, the archetypal FUor, shows that the CO absorption lines that are so broadened by the disk motion that they blend together \citep{Hartmann04}.  When unresolved, the absorption line equivalent widths are more consistent with a stellar spectrum than an FUor disk spectrum -- a conclusion that will be refuted in Section~\ref{sec:LineProfile}.

In the standard classification scheme, the spectra of FUors are considered as similar to giant or supergiant spectra because the FUor disks from which the spectra originate have low surface gravities \citep{Kenyon88, Welty92}.
In constrast, typical low-mass YSOs have surface gravities that are only slightly lower than main sequence dwarfs  \citep[e.g.][]{Cottaar14}.  
Since the strength of CO overtone absorption depends on surface gravity more than the Na and Ca lines, their relative equivalent widths usually indicate a luminosity class between dwarf and giant stars. Table~\ref{tab:EW} lists the equivalent widths of Na I doublet (2.206 and 2.209 $\mu$m), Ca I triplet (2.261, 2.263, and 2.266 $\mu$m), and an CO overtone band (2.292-2.320 $\mu$m), as measured with the {\tt\string splot} task of IRAF.  The equivalent width ratio of Na I $+$ Ca I to CO for IRAS 16316-1540 is consistent with those of Class I YSOs and dwarf stars and larger than those of FUors and giant stars (see Figure 9 in \citealt{Connelley18}; see also \citealt{Connelley10}).

\begin{table}[b!]
\centering
\label{tab:EW}
\caption{Equivalent widths of select absorption lines}
 \begin{tabular}{cccc} 
 \hline
 & & \multicolumn{2}{c}{EW in \AA} \\
 Object & Classification & (Ca + Na) & CO \\ 
 \hline
IRAS 16316-1540 & Class I & 4.3 & 11.3 \\ 
FU Ori & FUors & 6.5 & 32.9 \\
IRAS 18341-0113(N) & Class I & 8.1 & 27.8 \\
AA Tau & Class II & 5.4 & 16.3 \\ 
V830 Tau & Class III & 9.3 & 20.4 \\ 
 \hline
 \end{tabular}
\end{table}

\begin{figure*}[ht!]
    \includegraphics[width=0.99\textwidth]{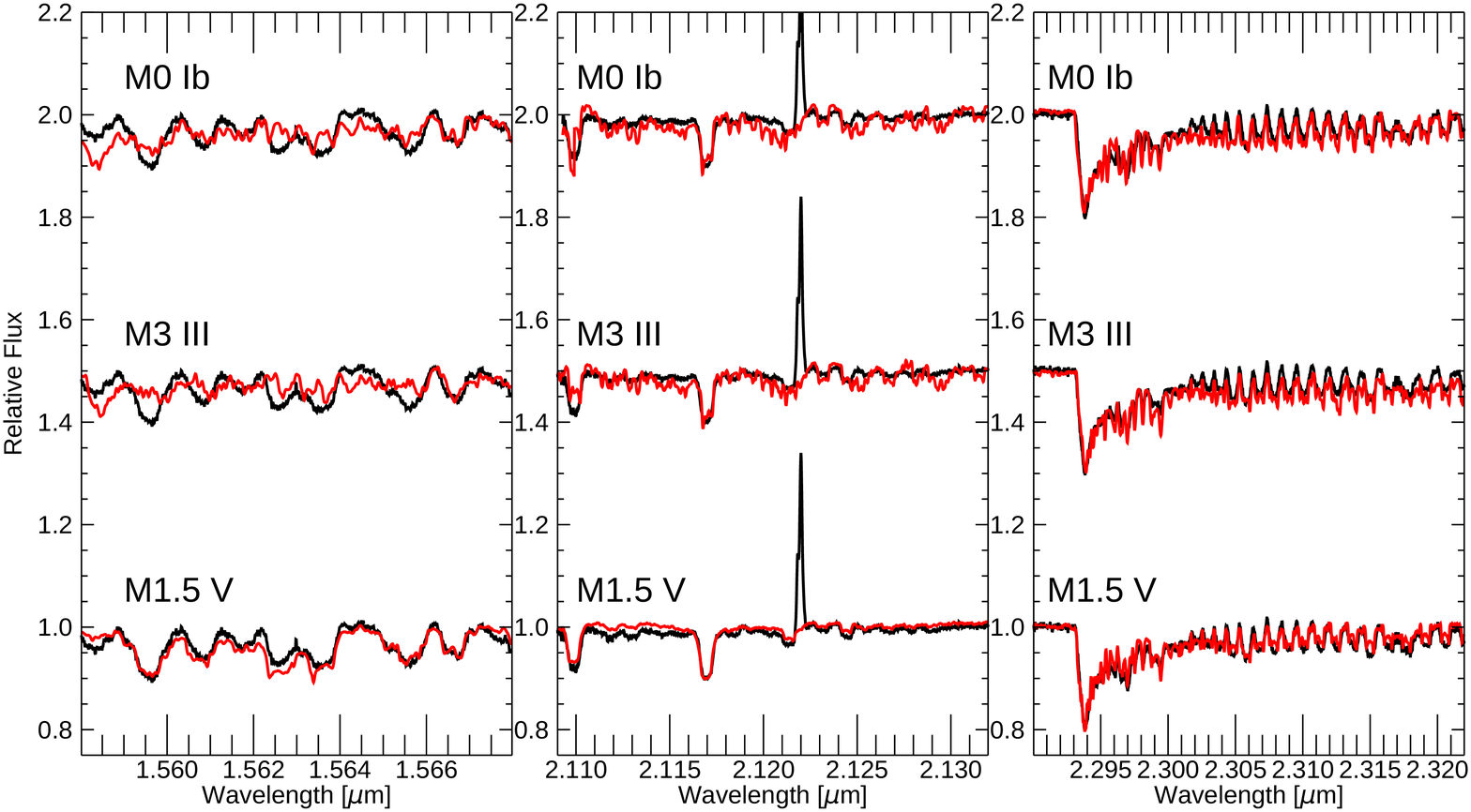}
    \caption{The comparison between the best-fit spectra of M1.5 dwarf, M3 giant, and M0 supergiant convolved with a disk rotation profile (see Section~\ref{sec:LineProfile}). The spectra of IRAS 16316-1540 and the standard stars are presented in black and red, respectively. The veiling is adjusted in each spectral region and for each possible template to best match the observed spectrum.  The position and profiles of the M1.5 V spectra are most consistent with IRAS 16316-1540. \label{fig:dwarftogiant}}
\end{figure*}

\begin{figure*}[ht!]
    \includegraphics[width=0.99\textwidth]{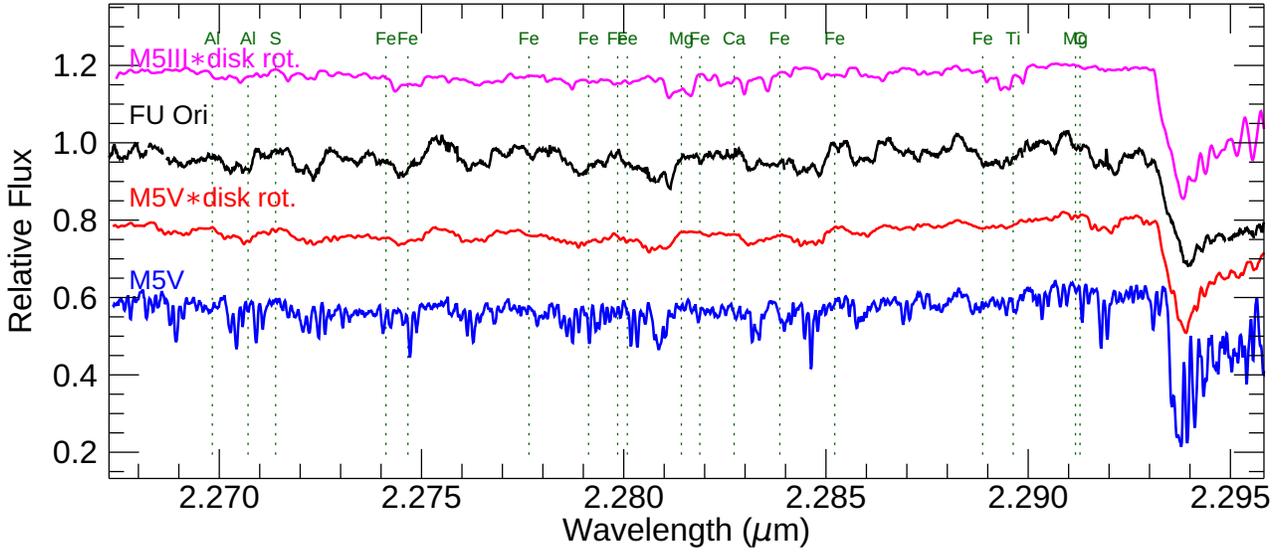}
    \caption{The spectra of FU Ori and the best-fit spectra of M5 dwarf and M5 giant. CO overtone band head and other spectral lines are shown in absorption. Black, red, and blue solid lines represent the spectra of FU Ori, the best-fit template spectra (M5 V) convolved with the disk rotation profile of 40 km s$^{-1}$, and the M5 V template spectra, respectively. For comparison, the M5 III template spectra convolved with the disk rotation profile of 45 km s$^{-1}$ are presented at the top (magenta). Green dotted lines show the positions of atomic lines of Arcturus \citep{Hinkle95}.
    \label{fig:FUOri}}
\end{figure*}

The approximate spectral type and gravity are further established by comparing the spectrum of IRAS 16316-1540 to spectra of M-dwarfs, giants, and supergiants (Figure~\ref{fig:dwarftogiant}), each convolved to a disk rotation profile and adjusted for veiling (see models in Section~\ref{sec:LineProfile}).
The overall features of the spectrum match the M-dwarf spectrum but do not match the M-giant and supergiant spectra, which disagrees with the fact that the FUors have the spectra of giant or supergiant. However, it does not mean that the dwarf spectra cannot account for the spectra of FUors. In the NIR, spectral features of FU Ori itself are similar to those of a M5 brown dwarf \citep[see Figure 11 in][]{Connelley18}. In our analysis of FU Ori, following that of IRAS 16316-1540, the K-band IGRINS spectrum is reasonably well-fitted with the template spectrum of an M5 dwarf \citep[see also][]{Hartmann87a,Kenyon88}, convolved with the disk rotation profile of $\sim$40 km s$^{-1}$ (Figure~\ref{fig:FUOri}). 
In low-resolution near-IR spectra, these many spectral lines are difficult to detect, so spectroscopic analyses are often based on the prominent CO overtone bands \citep[e.g.][]{Najita96}, which may lead to confused spectral classification.

In addition to the absorption spectra, several lines are seen in emission.  Br$\gamma$ 2.166 $\mu$m, an H I recombination line, is associated with the magnetospheric accretion, inner gaseous disks, disk winds, and outflows \citep{Muzerolle98a, Beck10, Salyk13, Tambovtseva16}. The double-peaked line profiles have peak separation of 179.2$\pm$15.8 km s$^{-1}$. The width, line profile variability, and self-absorption some epochs together indicate an origin in a magnetospheric flow \citep{Muzerolle98b,Kurosawa06,Eisner15}, unusual for an FUor.  Unlike the H$_2$ and [Fe II] lines, the lack of highly variable equivalent width of the Br$\gamma$ line indicates that the line emission is likely not spatially extended.

The H$_2$ and [\ion{Fe}{2}] lines typically trace shocked gas in outflows and jets \citep{Kumar99, Davis01, Davis03, Connelley10}. The double-peaked feature in H$_2$ emission with the peak separation of 26.2$\pm$1.4 km s$^{-1}$ indicates the bipolar outflow structure in IRAS 16316-1540 \citep{Arce06, Koyamatsu14}. The large variations in flux and modest variations in line profile are well explained by differences in seeing and slit size. For example, the lines are weakest during the observation at Gemini South, with $0\farcs6$ seeing, and strongest on nights with $>2^{\prime\prime}$ seeing at McDonald Observatory. Previous imaging demonstrated that the H$_2$ 2.122 $\mu$m emission extends over $\sim$20 arcseconds \citep{Kumar99}, so worse seeing will lead to stronger emission from nebulosity within the slit and extraction region, relative to the emission from the compact central source.

The broadband spectral shape may also change because of the differences in seeing between epochs.  The slope is reddest during the Gemini observation, when the seeing was a factor of 2 better than other epochs.  The $H$- and $K$-band images both have extended nebulosity, with the reddest $H$-$K$ color in the central region (see Figure 2 in \citealt{Mayama07}). If the nebulosity is scattered light, we would expect the equivalent widths of the line profiles to be preserved, as measured in our eight epochs.


\section{Line profiles and the origin of the spectra of IRAS 16316-1540}\label{sec:LineProfile}

In the previous section, we demonstrate that the spectrum of IRAS 16316-1540 appears consistent with M-dwarf photosphere, when analyzed through the presence and equivalent width of different spectral lines. However, the resolved line profiles tell a different story.  Their broad, boxy shapes are best explained from disk rotation and are inconsistent with stellar rotation.

\begin{figure*}[ht!]
\centering
\includegraphics[width=0.24\textwidth]{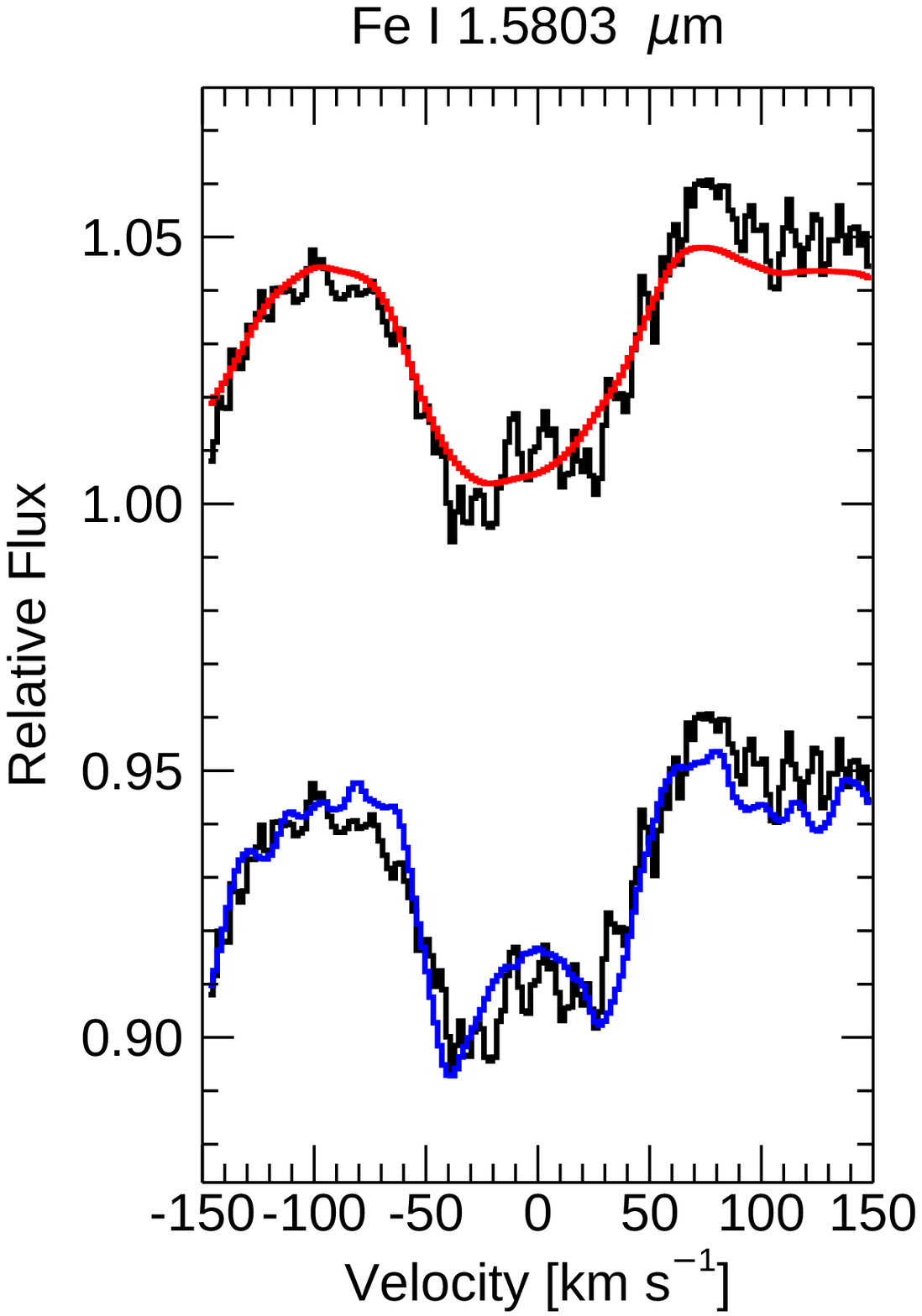}
\includegraphics[width=0.24\textwidth]{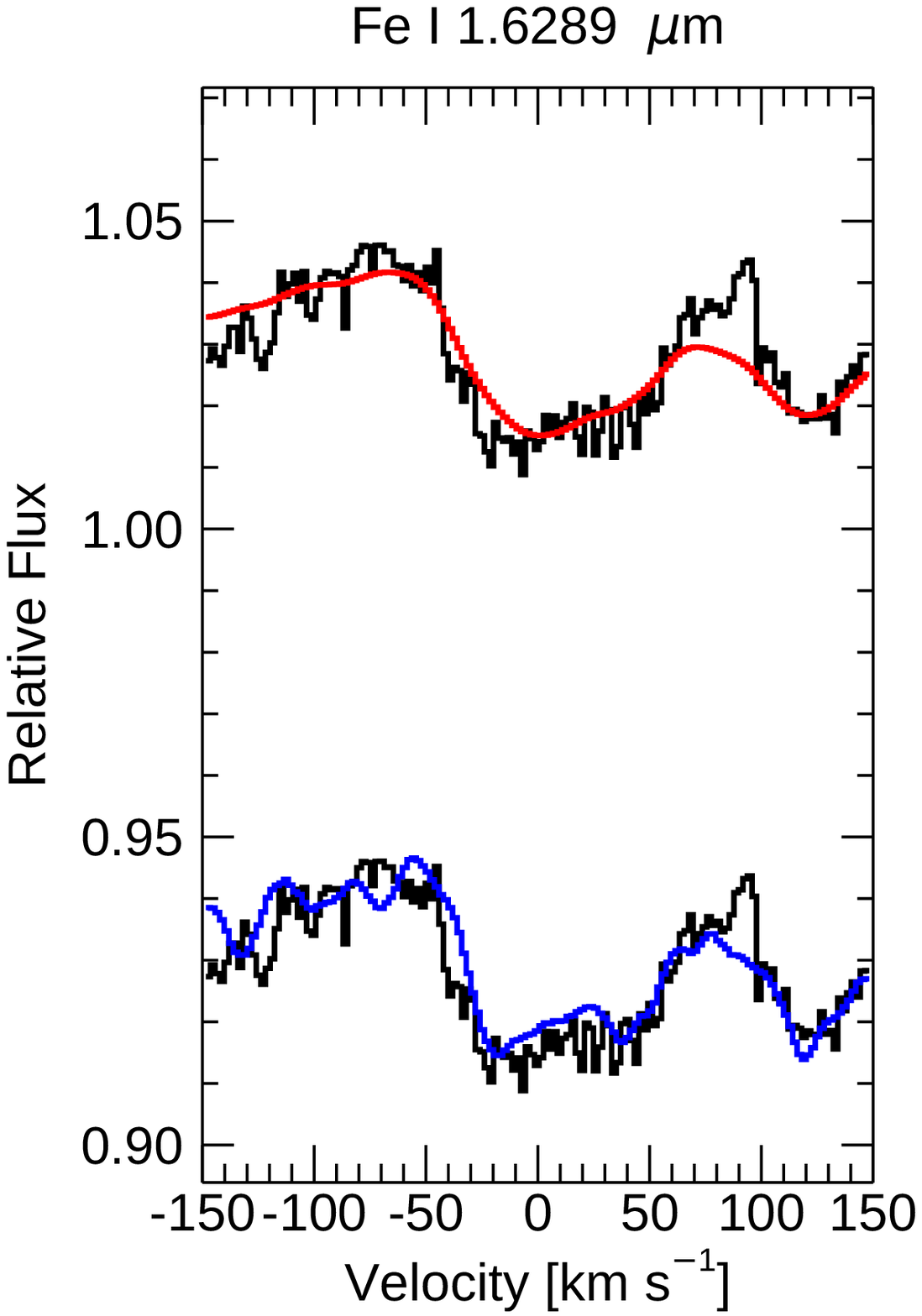}
\includegraphics[width=0.24\textwidth]{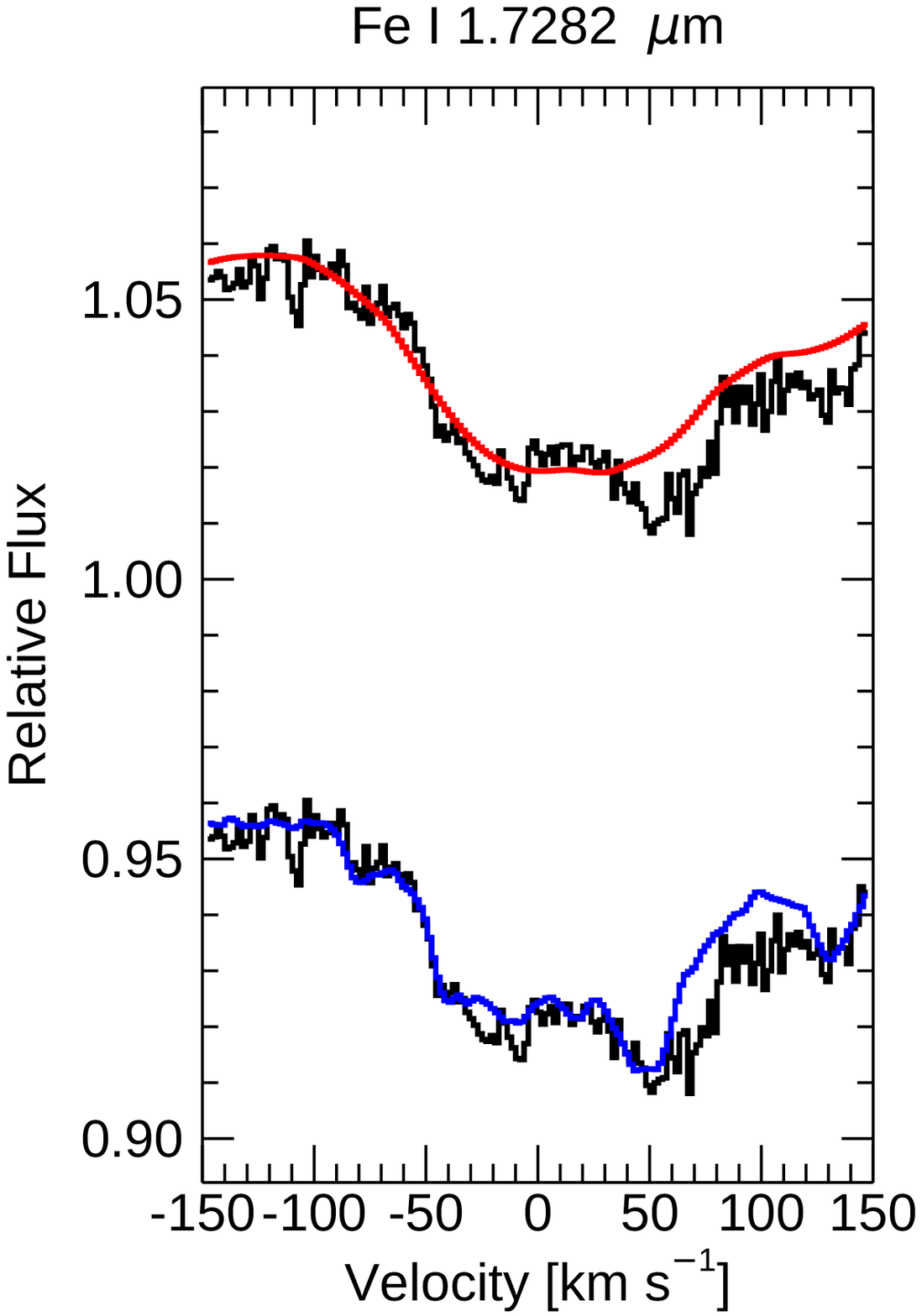}
\includegraphics[width=0.24\textwidth]{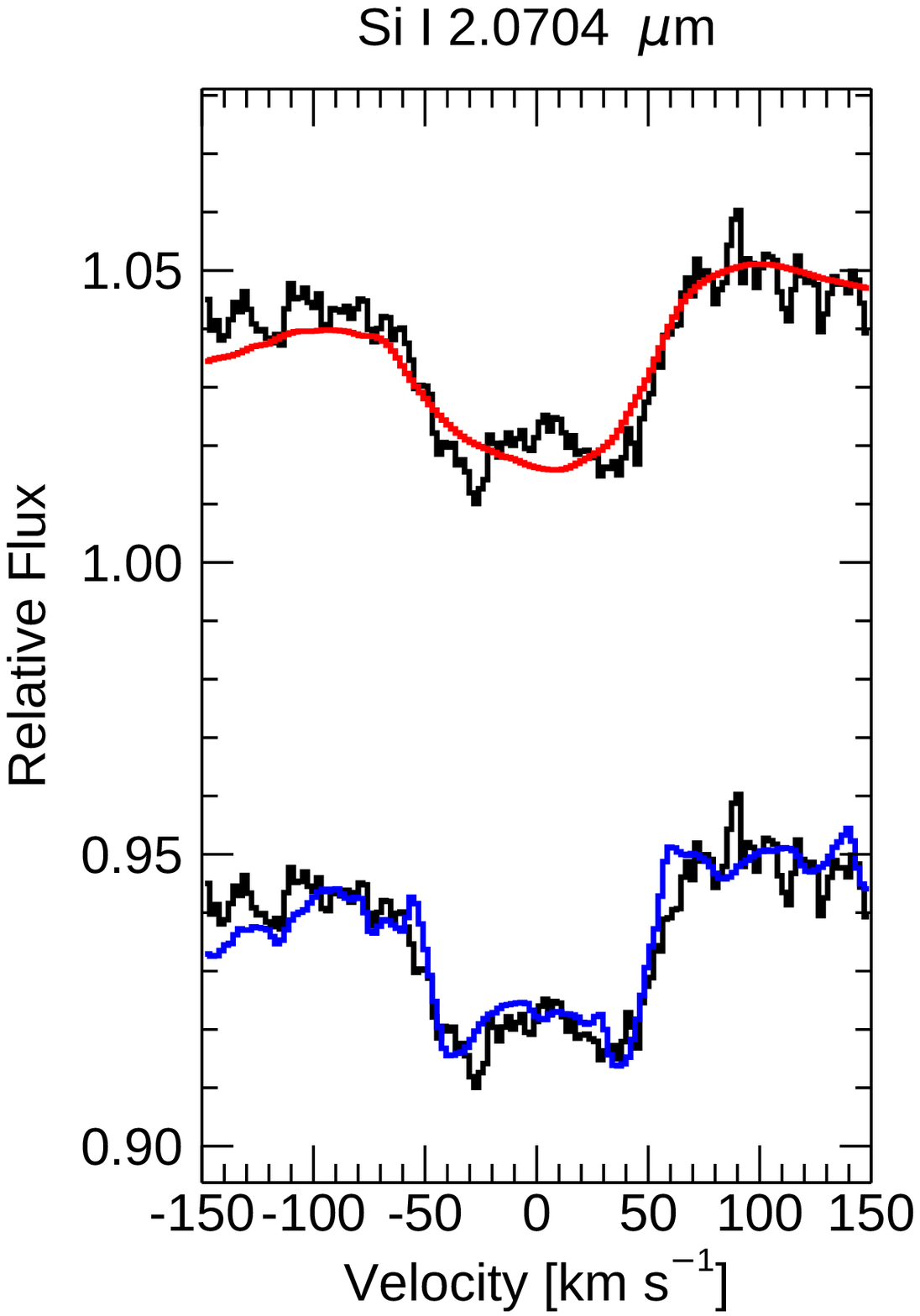}\\
\includegraphics[width=0.24\textwidth]{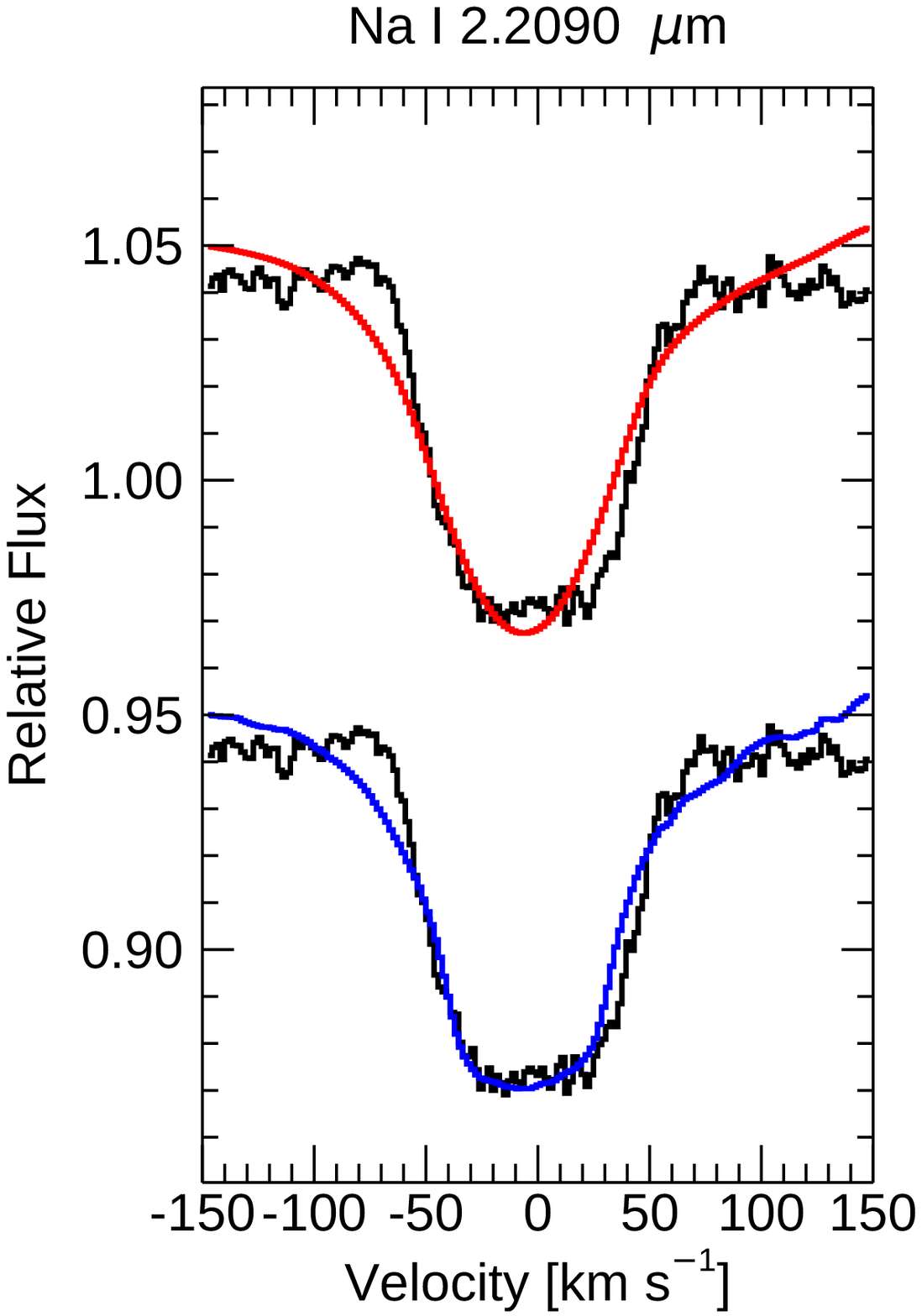}
\includegraphics[width=0.24\textwidth]{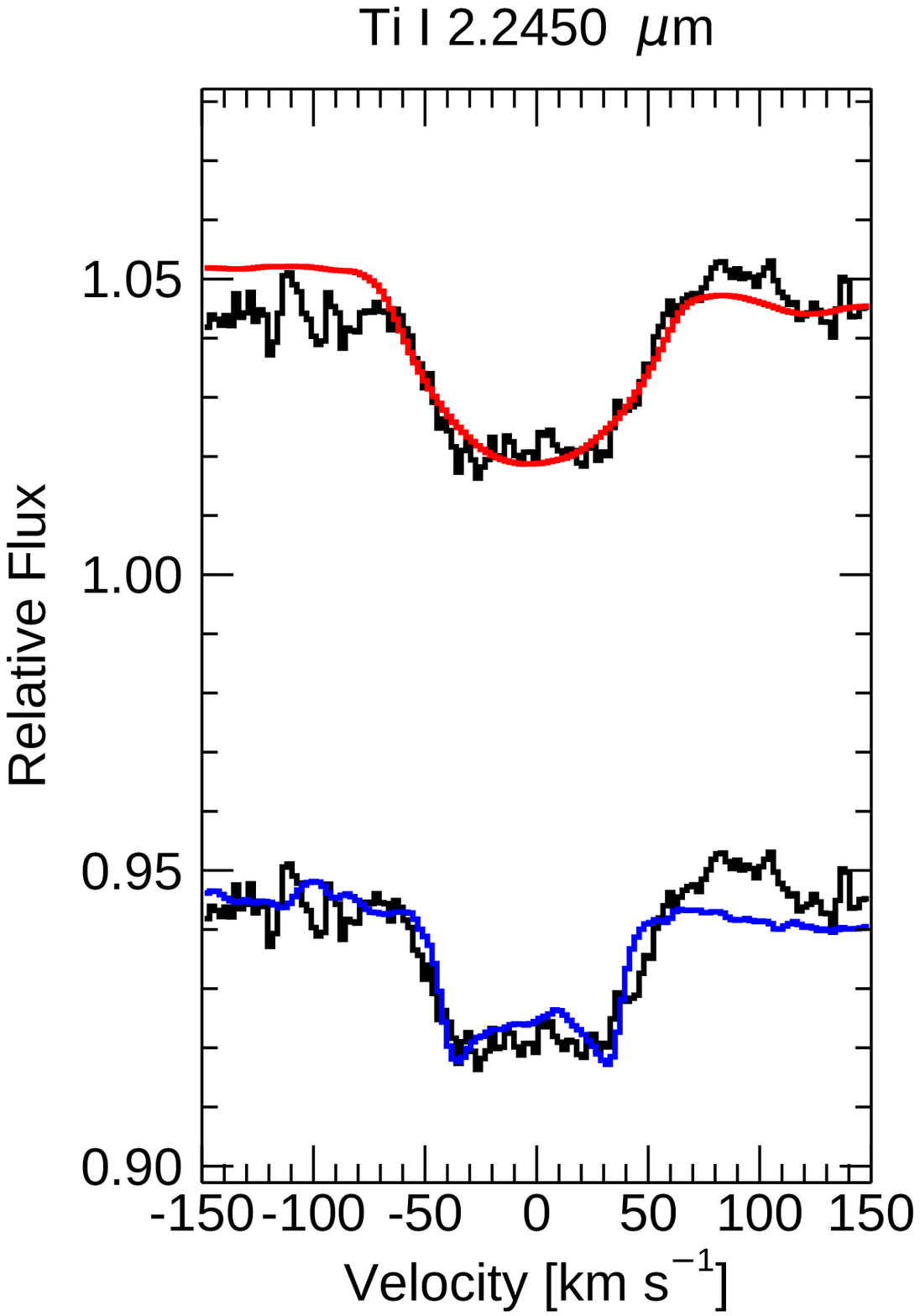}
\includegraphics[width=0.24\textwidth]{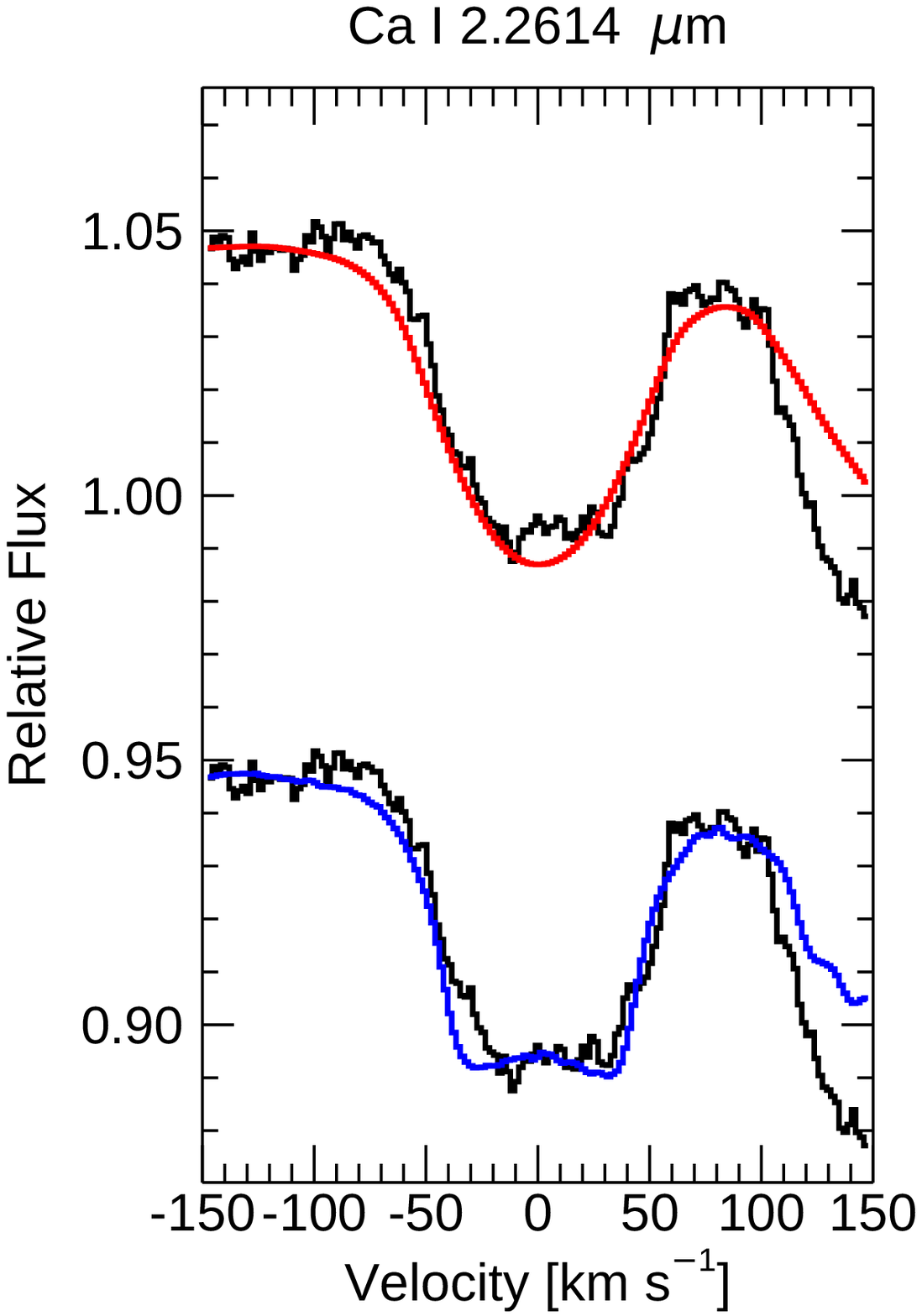}
\includegraphics[width=0.24\textwidth]{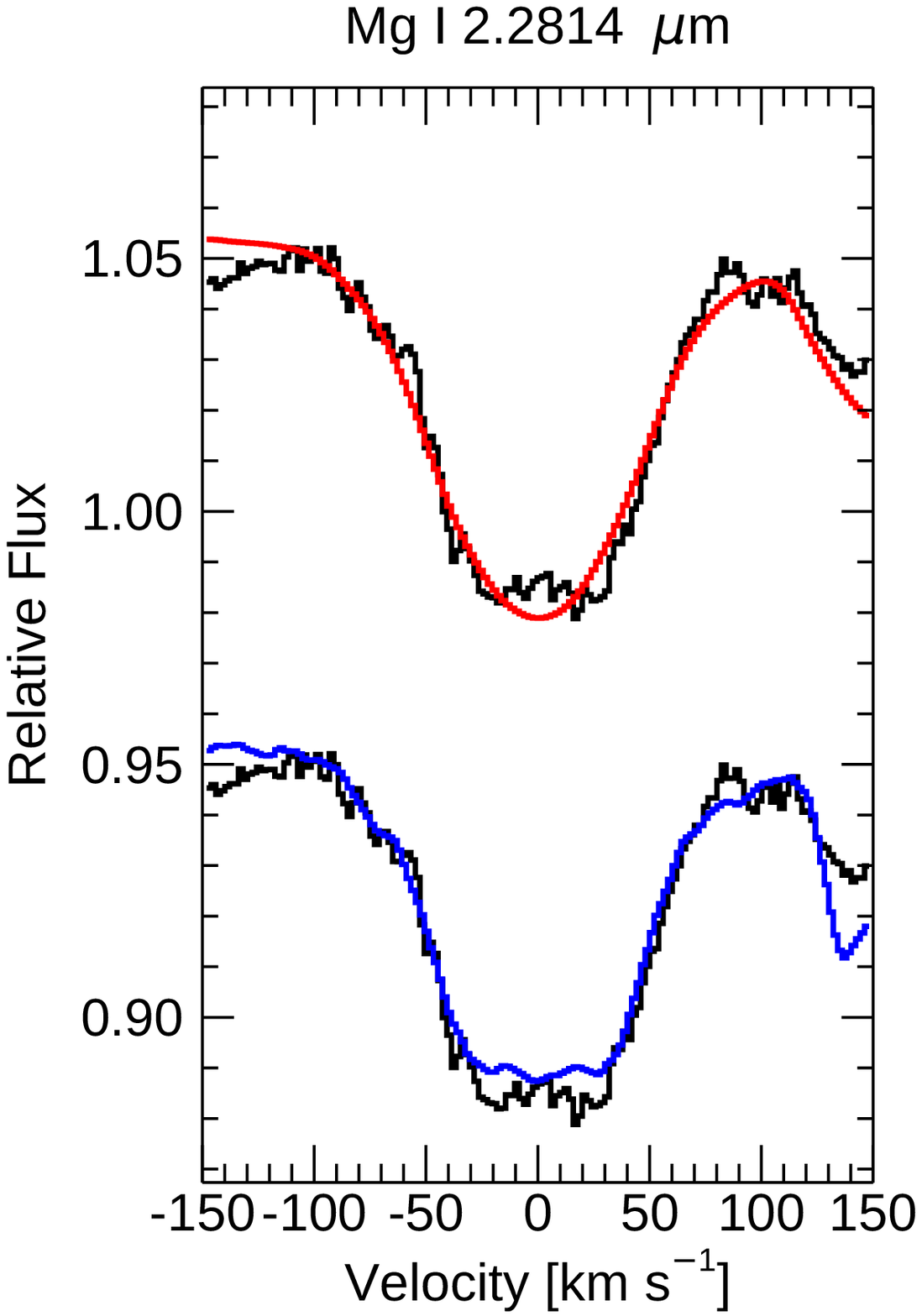}
\caption{Best-fit results for absorptions of Fe I 1.5803 $\mu$m, Fe I 1.6289 $\mu$m, and Fe I 1.7282 $\mu$m in H-band and Si I 2.0704 $\mu$m, Na I 2.2090 $\mu$m, Ti I 2.2450 $\mu$m, Ca I 2.2614 $\mu$m and Mg I 2.2814 $\mu$m in K-band spectra of IRAS 16316-1540. The template spectra of HD 36395 (M1.5 V) convolved with the stellar rotation profile of about 60.0 km $^{-1}$ (red) and the disk rotation profile of about 40.0 km $^{-1}$ (blue) are overlaid on the spectra of IRAS 16316-1540 (black) from top to bottom at each panel. \label{fig:convol}}
\end{figure*}

Figure~\ref{fig:convol} shows six atomic absorption lines of IRAS 16316-1540. All six have a broad, boxy line profile, with an average half width at half depth ($<$HWHD$>$) of 46.1$\pm$6.9 km s$^{-1}$, comparable to the HWHD of $\sim$36 km s$^{-1}$ of NIR absorption lines of FU Ori \citep{Zhu09a}. Spectra of FUors have double-peaked or boxy absorption line profiles in optical and NIR wavelengths, which are caused by the rotating disk with the mid-plane hotter than the atmosphere \citep{Hartmann96, Lee15, Park20}.    The rapid rotation of the protostellar photosphere also broadens absorption lines around a single peak \citep{Gray92}, and in principle could explain the width alone. The median rotational velocity is $\sim$38 km s$^{-1}$ (largest of $\sim$77 km s$^{-1}$) for Class I and flat-spectrum sources in nearby star-forming regions.
Two other potential broadening mechanisms, disk winds and companions \citep{Calvet93, Hartmann95}, are ruled out because the line profiles are centered at the radial velocity of the star \citep[$v_{{\rm lsr}}\sim0.5$ km s$^{-1}$;][]{Chen09} and do not vary with time.

We model the kinematic effects of the disk and stellar rotation in high-resolution photospheric absorption line profiles to distinguish between the disk and stellar photosphere scenarios for the formation of the NIR spectrum.
If the line profile traces a disk rotational profile, then the spectra originate from the accretion disk with a hot mid-plane, like FUors. On the other hand, if the line profile is the shape of a stellar rotational profile, then the continuum emission is produced in the photosphere of the protostar.

\subsection{Model line profiles}
 
{\it Accretion Disk:}  For absorption lines formed in the accretion disk, the line profiles will follow the disk rotation profile \citep{Hartmann85,Calvet93, Lee16}:
\begin{equation}
\phi(\Delta v) = \left[1-\left(\frac{\Delta v}{v_{\mathrm{max}}}\right)^2 \right]^{-1/2},
\end{equation}
where $v_{\rm max}$ is the maximum projected rotation velocity and $\Delta v$ is the velocity shift from the line center. We assume that the disk is optically thick, the vertical temperature gradient in the disk atmosphere can be treated as that of a stellar atmosphere, and the disk annuli are divided by a radial temperature distribution \citep{Kenyon88, Hartmann89, Welty92}. The NIR spectrum of the disk can then be represented by the spectrum of a standard star having an effective temperature corresponding to the temperature of a disk annulus \citep{Lee15, Park20}.  The lines are then convolved with the disk rotational profile.

{\it Stellar photosphere:} Absorption lines produced in a stellar photosphere have profiles described by a limb-darkened stellar rotation profile \citep{Gray92}: 
\begin{equation}
    G(\Delta v) = \frac{2(1-\epsilon)[1-(\Delta v/v_{\mathrm{max}})^2]^{1/2} + \frac{\pi\epsilon}{2}[1-(\Delta v/v_{\mathrm{max}})^2]}{\pi v_{\mathrm{max}}(1-\epsilon/3)},
\end{equation}
where $\epsilon$ is the limb-darkening coefficient and assumed as the typical value of 0.6 for the photospheric lines \citep{Doppmann05, Covey05}.  The lines are then convolved with the stellar rotational profile.

For both the disk and stellar photosphere, the model spectra are created by using template spectra from the IGRINS spectral library\footnote{http://starformation.khu.ac.kr/IGRINS\_spectral\_library.htm} \citep{Park18}, which covers spectral types from O to M and luminosity classes from I to V. We took the G- to M-type spectra of standard stars from the IGRINS spectral library as template spectra of the protostellar disk and photosphere.

{\it Veiling:}  Whether from a photosphere or from a disk, the convolved template spectra have deeper lines than those of IRAS 16316-1540. To match the EW of template spectra to IRAS 16316-1540, we introduce a veiling factor, defined as
\begin{equation}\label{eq:veiling}
    r = \frac{\mathrm{EW_{int}}}{\mathrm{EW_{obs}}}-1,
\end{equation}
where EW$_{\rm int}$ and EW$_{\rm obs}$ are the intrinsic (template spectra) and observed (IRAS 16316-1540) equivalent widths, respectively.  Veiling in the NIR is typically interpreted as excess continuum emission from warm dust in the inner disk (or for the optical, from accretion) that reduces the depth of a photospheric line \citep[e.g.][]{Johns-Krull01, Vacca11}. For a spectrum where the photosphere is the disk itself, veiling is less well defined physically, though a similar concept may apply. In both cases, the veiling is a numerical parameter used so that the model and observed line profiles have the same depth.  Veiling is applied to the convolved spectra in steps of 0.1  from 0.0 to 10.0.

\subsection{Comparing the template spectra to observed line profiles}

We select 61 spectral regions in the IRAS 16316-1540 spectra that contain well-isolated lines to compare the line profile with those of model spectra.
We also include CO overtone band regions, despite crowding, because the CO bands are the prominent feature in the spectra of FUors and typical YSOs. We compare the fitting parameters obtained from the isolated line regions to those obtained from the CO bands to examine the consistency between the relatively weak spectral regions and the prominent CO spectra. 

Every template spectrum in the IGRINS spectral library is convolved with disk or stellar rotational profiles with widths from 10 to 80 km s$^{-1}$ in steps of 2.5 km s$^{-1}$, shifted to match the line center of the IRAS 16316-1540 spectrum, and normalized to the observed continuum level.  The best-fit parameters from $\chi^2$ minimization then, together with the visual inspection, lead to spectral type and luminosity class from the best fit of the template spectra, along with the associated rotational velocity and veiling factor. 

Among template spectra, based on the $\chi^2$ values and visual inspection, we determined the M1.5 V template spectra of HD 36395 as best-fit results for both disk and stellar rotation profiles; both disk and stellar best-fit results have the mean $\chi^2$ of $\sim$1.0 over the 8 lines in Figure~\ref{fig:convol}. The mean rotational velocity and veiling factor\footnote{The listed uncertainties are standard deviations from the 61 different regions.} are 41$\pm$5 km s$^{-1}$ and 0.54$\pm$0.45 for the disk rotation profile, and 60$\pm$8 km s$^{-1}$ and 0.45$\pm$0.45 for the stellar rotational profile. 

Figure~\ref{fig:convol} shows the best-fit spectra to the IRAS 16316-1540 spectra for the disk and stellar rotation profiles.
The disk rotational profile provides a better fit to the boxy, double-peaked spectral shape at the line center regime than the rounded profile from stellar rotation. Based on this result, we suggest that the spectral characteristics of IRAS 16316-1540 originate in the disk.

\begin{figure*}[ht!]
\centering
\includegraphics[width=0.45\textwidth]{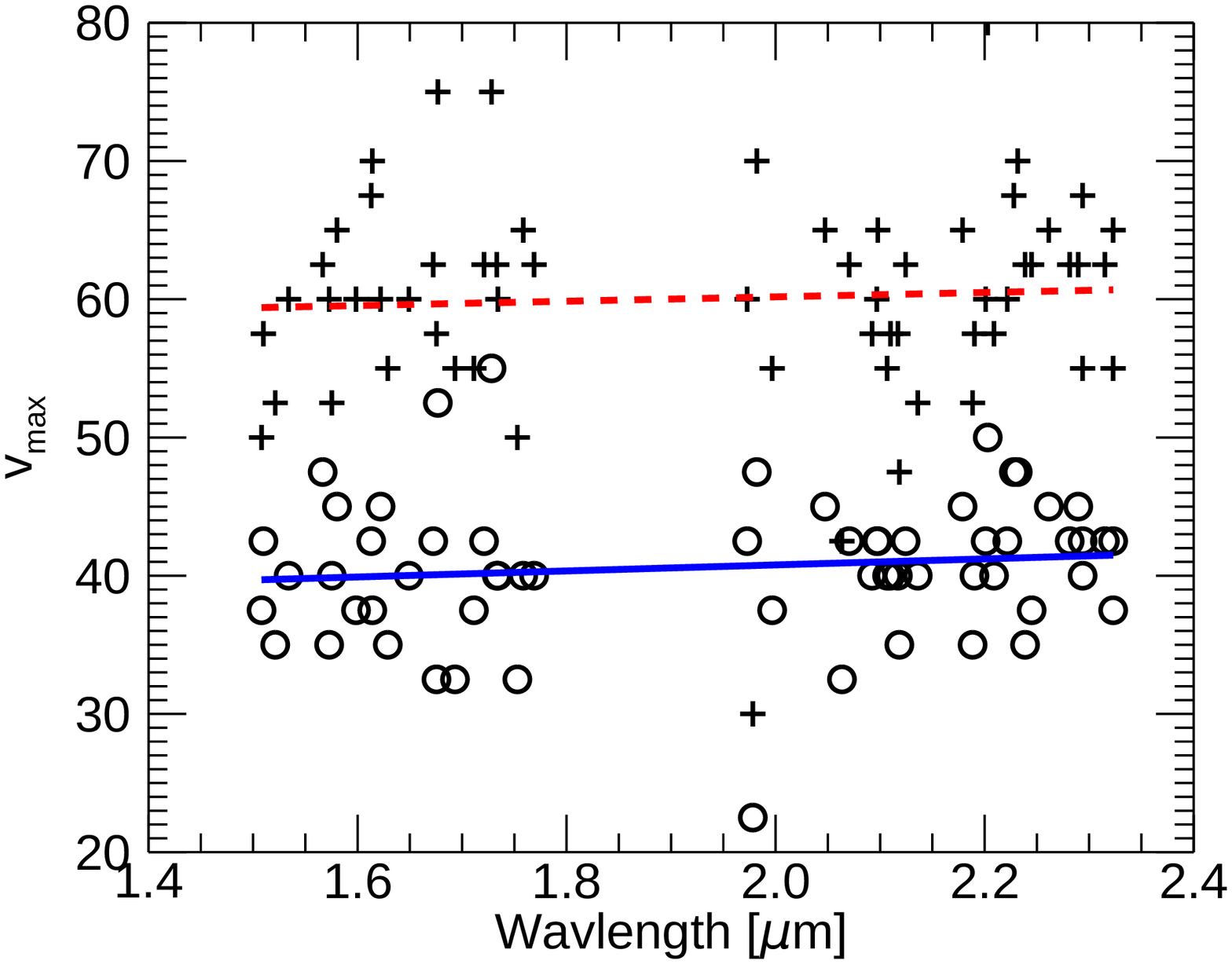}
\includegraphics[width=0.45\textwidth]{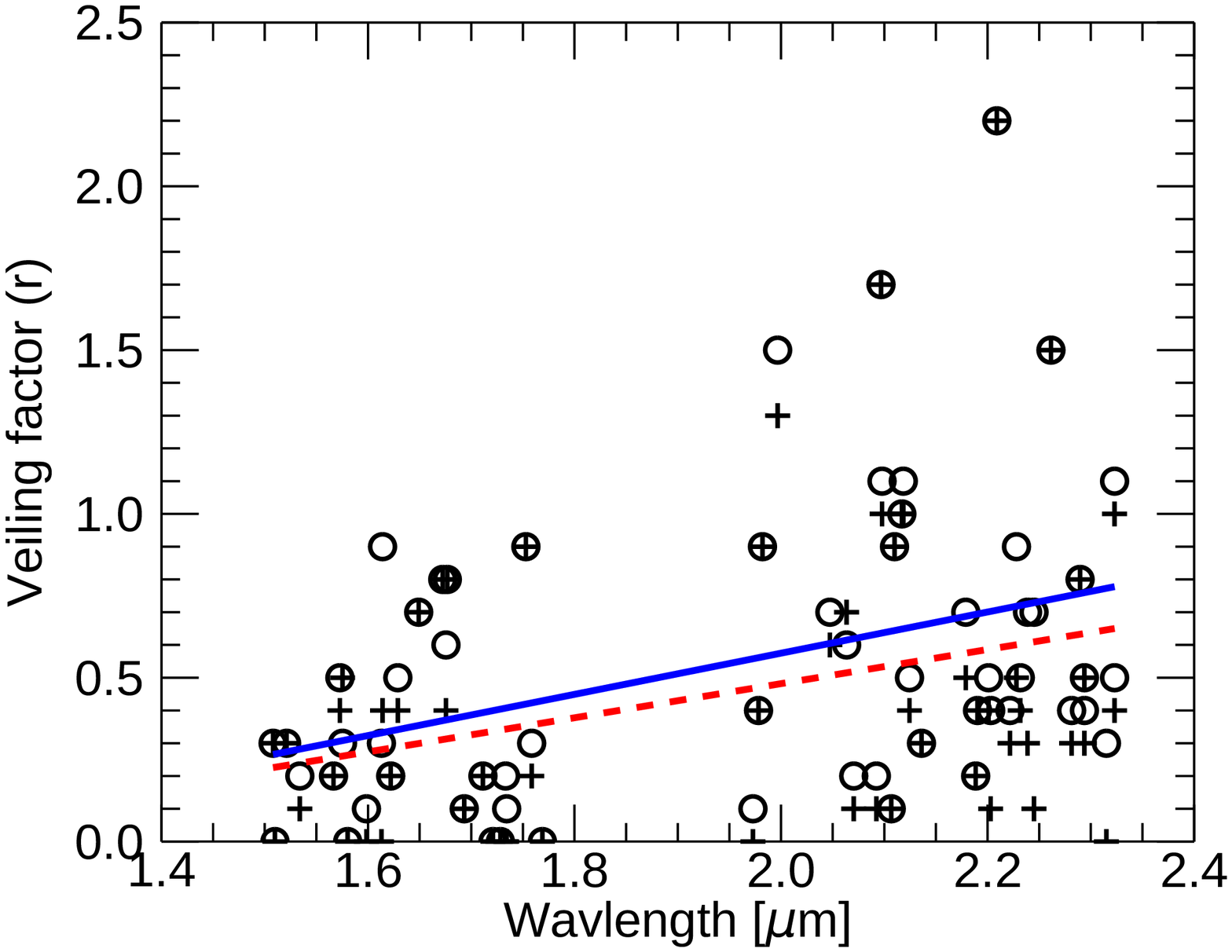}
\caption{Dependence of rotational velocity (left panel) and veiling factor (right panel) on wavelength. The plus signs indicate the estimated values for the stellar rotation profile, and the open circles denote values determined for the disk rotation profile. The red dashed and blue solid lines show linear relations for the stellar and disk rotation profile, respectively. \label{fig:veiling&vmax}}
\end{figure*}

The projected rotational velocities are constant with wavelength within our NIR spectra (Figure~\ref{fig:veiling&vmax}). In a Keplerian disk model, the rotational velocity of the disk is expected to decrease with wavelength because the warmer regions close to the protostar rotate faster than cooler outer regions \citep{Hartmann87a, Hartmann87b, Welty90}. FUor disks show this increase in line width towards shorter wavelengths, but only when comparing optical to the infrared \citep{Petrov08, Zhu09a, Lee15, Park20}. The IGRINS spectral coverage alone is insufficient to measure any differences in line widths versus wavelength.

The NIR absorption lines can be veiled by the thermal dust emission attributed to the accretion disk and envelope \citep{Calvet97, Greene02, Nisini05}. The trend of increase in the veiling factor with wavelength is explained by the thermal dust emission (Figure~\ref{fig:veiling&vmax}). The infall model of \citet{Calvet97} predicts the veiling in K-band is larger than in H-band. The veiling factors measured with each spectral lines in the disk and stellar model spectra have means\footnote{The listed uncertainties are the standard deviation of the measurements.} of 0.54$\pm$0.45 for the disk model and 0.45$\pm$0.45 for the stellar model \citep[consistent with veiling estimated in the low-resolution spectra of][]{Connelley10}. Since the veiling factor is measured by fitting each line independently, the large scatter in the veiling factor of nearby absorption lines is unexpected, given that the 1 sigma errors for the veiling factors are less than 0.1 in most cases.  The large scatter in the veiling factor might reflect the intrinsic difference between the disk atmosphere with a vertical and radial temperature gradient and the template stellar atmosphere with a single effective temperature.  Rather than just using the template spectra, a detailed disk model is needed for future study to evaluate any NIR veiling in a disk spectra.

\section{Discussion}\label{sec:Discuss}

We have shown that the high-resolution NIR spectra of IRAS 16316-1540 have boxy, double-peaked absorption lines, which are matched with the M dwarf template spectra convolved with the disk rotation profiles of 41$\pm$5 km s$^{-1}$. For protostars with high enough accretion rates, the disk mid-plane is heated by viscous dissipation \citep[see reviews by][]{Hartmann96, Audard14}. The projected disk rotation velocity corresponds to the disk annuli located at $\sim$0.1 AU from the central star, for an assumed stellar mass of 0.3 $\rm M_{\sun}$ and an estimated disk inclination of 60$^{\circ}$ \citep{Weintraub94}.
The radiation from the disk mid-plane is absorbed by relatively cool disk atmospheres, and then the rotation of the disk broadens the absorption line making the double-peaked or boxy profile. 

The viscous heating of the disk requires high accretion rates, typically those found only for FUor objects \citep[e.g., see discussions in][]{Hartmann16}.   Since we do not detect the burst itself, we classify IRAS 16316-1540 as a candidate FUor (or an FUor-like object, \citealt{Greene08}). Moreover, our analysis demonstrates that some viscously-heated disks (FUor candidates) may have been misclassified as photospheres. The bias is the opposite of the concern demonstrated by \citet{Connelley18} that some protostars with mid- to late-M spectral types may be misclassified as FUors because of similar depths in absorption features.

Discrepancies between the template and observed spectra, as indicated by the large scatter in veiling factors measured from each line, point to a deficiency in the use of stellar photospheres as templates.  It may be due to that the fundamental physical differences between the template stellar atmosphere and the disk. Although the dwarf spectral features are seen in the observed spectra, the disk atmosphere will be different than the stellar atmosphere, including having a lower surface gravity.  Further investigations with disk models and observations of viscously heated disks are necessary to overcome the limitation of this work.

Nevertheless, we discuss a property that appears in both dwarfs and FUor disks in order to suggest a possibility of what makes the disk spectra similar to the dwarfs. The dwarf spectral features in FUor/FUor-like objects might be related to the convective flows in the viscously heated disk. As the material accumulates in the inner region of the disk, the surface density increases, and the mid-plane is heated by the viscous dissipation. At this point, the disk becomes optically thick in the infrared as the opacity increases, which renders the inner region convectively unstable \citep{Zhu09b, Pavlyuchenkov20, Zhu20}. Similarly, late-type dwarfs have a fully convective interior, and the materials in the star are well mixed. Therefore, the abundance in the stellar atmosphere shows the abundance in the core \citep{Allard97}. In that way, the convection in the accretion disk enables the molecular species formed in the disk mid-plane to be transported to upper layers. Thus, the disk spectra might have somewhat similar spectral features to M-dwarfs.

Despite our interpretation of IRAS 16316-1540 as an FUor-like object, the spectra has some unusual properties when compared to the broader FUor class. The line depths are more similar to a dwarf than other FUor objects, which had led to the prior classification of the spectrum as a stellar photosphere. The bolometric luminosity is a factor of 10 lower than the bonafide, historical set of FUors, although some FUor-like objects have similar luminosities \citep{Connelley18}. The IGRINS spectrum shows Br$\gamma$ emission \citep[see also H-line detections by][]{Connelley10,Salyk13}, which suggests magnetospheric accretion, a process not expected at such high accretion rates. Most FUors show Br$\gamma$ in absorption rather than emission, although in the declining phase, Br$\gamma$ can appear in emission \citep{Kun19}. The warm inner disk is detected in CO fundamental emission in the $M$-band \citep{Herczeg11}, despite the absorption in CO overtone bands described here. These characteristics of IRAS 16316-1540 add to the rich fabric of protostellar outbursts, which in many cases do not fit neatly into our classification system \citep{ContrerasPena17}, and should help to guide future modeling efforts.

\subsection{Ice Features as supporting evidence of an accretion burst in IRAS 16316-1540}\label{subsec:addobs}

The existence of an accretion burst should be imprinted in the ice feature of the protostar, though we cannot pinpoint the moment. The \textit{Spitzer} IRS spectrum shows a double-peaked absorption feature of the pure CO$_2$ ice component \citep{Pontoppidan08}, which indicates that the burst state of IRAS 16316-1540 has existed. The chemical distribution can be used as a diagnostic of the accretion burst \citep{Lee07}, since the accretion burst changes the physical and chemical structures of the disk and envelope \citep{Visser15}. In dense and cold environments, molecules such as H$_2$O and CO are easily frozen onto dust grains and depleted from the gas. These simple ice species are incorporated into more complex ones, such as CO$_2$ and CH$_3$OH, on cold grain surfaces. The CO$_2$ ice, for example, is mixed with CO and H$_2$O ices in grain surfaces. However, if dust grains are heated above 20 K, CO$_2$ ices are isolated by the sublimation of other ice species.  This process is irreversible, since the binding energy between CO$_2$ molecules is greater than between CO$_2$ and other species. Therefore, the double-peaked absorption in the pure CO$_2$ ice feature at 15.2 $\mu$m in the mid-IR spectrum can be used to trace the high-luminosity state of a protostellar system in the past \citep{Kim12}, including IRAS 16316-1540. 

\subsection{Accretion burst by disk precession?}\label{subsec:precession}

IRAS 16316-1540 is associated with complex molecular outflows. On a spatial scale of about $500 \arcsec$, the blue-shifted outflow extending to the southeast sweeps away the material and forms a U-shaped shell, while the red-shifted outflow is driven to the northwest \citep{Bence98, Lee02, Lee05}. In the innermost region near the source ($\lesssim$ 30$\arcsec$), the blue-shifted outflow is driven southward while the red-shifted outflow is driven northeastward \citep{Lee00, Arce06, Mayama07, Koyamatsu14}. The outflow axes are misaligned between large ($\sim 500 \arcsec$) and small ($\lesssim$ 30$\arcsec$) spatial scales, which could be interpreted as a result of a disk precession \citep{Lee00, Mayama07, LeeJE07}. 

Anisotropic accretion by infalling clumps can explain the accretion burst in the protostellar system having complex outflow and envelope structures. The infalling knots that formed far from the central star are likely to have a different angular momentum compared to the protostar-disk system. When the condensed mass collides with the disk, it leads to a change in the angular momentum of the disk, and the perturbed disk would be gravitationally unstable. Eventually, such anisotropic accretion can cause the accretion burst \citep{Kuffmeier18}.

Misaligned features in outflows were also found in the FUor object V1647 Ori, an embedded protostar that has had multiple eruptions \citep{Reipurth04, Aspin06, Aspin09}.  
Based on misaligned axes between the approaching and receding outflows towards the northwest and the south, \citet{Principe18} suggested that the outbursts are caused by infalling clumps onto the disk. IRAS 16316-1540 is also surrounded by the infalling knots in a dissipating envelope and has a misaligned outflow structure \citep{Schild89, Mayama07, Koyamatsu14}. Therefore, if the infalling knots collided onto the disk, it can cause the re-orientation of outflow direction and a significant increase of mass accretion onto the central object of IRAS 16316-1540.

\section{Summary and Conclusion}\label{sec:final}

In this paper, we used  IGIRNS spectra of the embedded protostar IRAS 16316-1540 to demonstrate that the near-IR continuum emission is produced in a viscously-heated disk. 
We compared  the spectra of IRAS 16316-1540 with the template spectra convolved with the rotation profiles of the disk or stellar photosphere in order to investigate the origin of the spectra, with the following conclusions:

(i) The photospheric line profiles are double-peaked or boxy, consistent with the template spectra of M1.5 V convolved with a disk rotation profile of 41$\pm$5 km s$^{-1}$ and inconsistent with a stellar photopshere. Therefore, we conclude that the NIR emission of IRAS 16316-1540 originates in the disk that heated by an accretion burst, like in FUors. 

(ii) The M dwarf spectral feature in the NIR spectra of IRAS 16316-1540 and FU Ori conflicts with that the disk models predict a giant/supergiant spectrum in the spectra of FUors. There is a limitation in fitting the disk spectrum with a template stellar spectrum because the disk has a range of temperature, may have abundance anomalies, and pressures are very different. The dwarf spectral features in the FUors spectra might be related to the convective flow in the accretion disk, but still not clear. To understand the inconsistency, more samples of high-resolution spectra of FUor and FUor candidates are needed, and also a disk model is needed to explain the spectral features.

(iii) The pure CO$_2$ feature in the mid-IR spectrum supports the idea that a burst state existed in IRAS 16316-1540. The change in the outflow direction with the distance from IRAS 16316-1540 could result from the disk precession caused by the impact of infalling clumps onto the disk, which might trigger the accretion burst with disk instability.

(iv) In the growth of the star, significant unresolved issues still remain, including how protostars evolve at the earliest stage, how frequently accretion bursts occur in embedded protostars, and how long these burst lasts. These uncertainties are mainly due to the scarcity of outburst samples in the early protostellar stage. High-resolution spectroscopy toward the embedded protostar, IRAS 16316-1540, reveals the line profiles expected from the disk heated by an accretion burst, despite previous low-resolution spectra that indicated the near-IR emission was produced in a stellar photosphere.  Since high-resolution spectroscopy is expensive, some ongoing bursts may have been overlooked.
An extension from this high resolution spectroscopic observation to more embedded protostars will help to evaluate the importance of episodic accretion process in the early evolutionary protostellar phase.

(v) The physical and chemical structure of the disk affected by the accretion burst in the scale of tens to hundreds of AU has been unexplored in detail. Heating from the accretion flow sublimates molecular ices from the dust grain surfaces within so called snow line. When the accretion burst occurs, the snow line extends to much larger radii \citep{Lee19}. Through high-angular resolution imaging observations towards the protostellar disk, we can directly study the disk structure and the influence of the accretion burst on the disk. ALMA, with unprecedented angular resolution and sensitivity, will enable us to study in more detail the physical and chemical properties in the disk of IRAS 16316-1540. Therefore, combining the NIR and sub-mm observations will bring a holistic comprehension of this embedded protostellar disk.

\acknowledgements
This work was supported by the National Research Foundation of Korea (NRF) grant funded by the Korea government (MSIT) (grant number 2021R1A2C1011718). GJH is supported by research grant 11773002 from the National Science Foundation of China.  This work used the Immersion Grating Infrared Spectrometer (IGRINS) that was developed under a collaboration between the University of Texas at Austin and the Korea Astronomy and Space Science Institute (KASI) with the financial support of the Mt. Cuba Astronomical Foundation, of the US National Science Foundation under grants AST-1229522 and AST-1702267, of the McDonald Observatory of the University of Texas at Austin, of the Korean GMT Project of KASI, and Gemini Observatory.

\acknowledgements
This work used the softwares as follows 
\software{IRAF \citep{Tody86, Tody93}, 
IGRINS pipeline \citep{Lee17}
}

\bibliographystyle{aasjournal.bst}
\bibliography{IRAS16316_draft_YSY.bib}

\end{document}